\documentstyle[12pt,epsf,epsfig]{article}

\def\bm#1        {{\bf #1}}

\def\GeVc2       {{\rm GeV}/c^2}

\def\lapproxeq   {\lower .7ex\hbox{$\;\stackrel{\textstyle
                  <}{\sim}\;$}}
\def\gapproxeq   {\lower .7ex\hbox{$\;\stackrel{\textstyle
                  >}{\sim}\;$}}
\def\beq         {\begin{equation}}
\def\eeq         {\end{equation}}
\def\shift       {\rule[-3mm]{0mm}{8mm}}

\begin{document}

\begin{titlepage}

\vspace*{0.1cm}   
\begin{flushright}  
DTP/96/86        \\
hep-ph/9609526  \\
September  1996  \\   
\end{flushright}   
\vskip   1.2cm   

\begin{center}
{\Large\bf   
Testing a Model with additional Vector Fermions \\[2mm]  
at the LEP2 Collider } 
\vskip 1.cm 
{\large  M.~Heyssler\footnote{Electronic ddress: M.M.Heyssler@durham.ac.uk}
and V.C.~Spanos\footnote{Electronic address: V.C.Spanos@durham.ac.uk}}  
\vskip .3cm
{\it  Department  of Physics,  University of Durham, \\ 
Durham DH1 3LE, England }\\

\vskip 1cm

\end{center}

\begin{abstract}

Our  aim is to  test a  model  recently  presented,  motivated  by the
reported  $R_b$ and  $R_c$  ``crisis",  which  contains  extra  vector
fermions.  We suggest an  alternative  indirect  test of the  possible
existence  of new heavy quark  flavours  at the LEP2  collider,  which
turns out to give the clearest signal.  We calculate  $q\bar{q}$ cross
sections within this framework,  including one loop  corrections,  and
find   measurable   differences   compared  to  the   Standard   Model
predictions.

\end{abstract}

\vfill 

\end{titlepage}

\newpage

\section{Introduction}

Motivated by the  so--called  $R_b, R_c$ crisis that arose  during the
last couple of years,  several  models have been  invoked,  to explain
this puzzle using different kinds of extensions to the Standard  Model
(SM).   Besides   (leptophobic)   $Z'$   models   and   supersymmetric
corrections,  there were several  attempts  which  introduce new heavy
vector  fermions  \cite{VFModels96,Chang96}.  Even  though  the latest
publications by the LEP collaborations  \cite{Blondel96}  show a value
of $R_c = 0.1715\pm  0.0056$, which hence lies within the  predictions
of the SM, and the  $R_b$  value is  reported  to be $R_b =  0.2178\pm
0.0011$, yet in excess by only  $1.9\sigma$  of the SM, the  suggested
models  cannot be entirely  excluded and it seems  worthwhile  at this
stage to find an alternative test of these models, such that a picture
of   consistency   may  (or  may  not)  emerge.  There  are  now  many
predictions on how to test the $Z'$ model  \cite{Heyssler96} in a more
general framework  independently  from $R_b$, $R_c$  measurements, but
there is still lacking  information  about the  testability of the new
vector fermion models beyond their genuine motivation.

In this paper we are considering the model  suggested by Chang {\it et
al.}  \cite{Chang96},  which  introduces  a  new  heavy  vector  quark
triplet  $X$ mixing  with the quarks of the SM.  As a  consequence  of
this mixing, there are effects to be expected, showing up in different
processes  as stated  below.  Adopting  especially  this model, we can
simultaneously  solve for possible excesses of $R_b$ as well as $R_c$,
without introducing anomalies.

The possible tests for this model could be:

\begin{description}
\item[{(a)}]
{the calculation of, e.g., heavy flavour  production,  analogue to the
$Z'$ at hadron  colliders, to judge,  whether a significant  change in
the cross section can be observed,}
\item[{(b)}]
{the study of  flavour--changing  neutral  current  (FCNC) effects in,
e.g., decay modes of some mesons like $K^0_L$ or $B^0$,}
\item[{(c)}]
{low energy physics experiments, e.g., $\nu N$ scattering,}
\item[{(d)}]
{quark--antiquark  production at $e^+e^-$ colliders, especially at the
CERN LEP2 collider.}
\end{description}

An  analogous  treatment to the $Z'$ model like in (a) is not possible
because  the new vector  and  axial--vector  couplings  induced by the
mixing  with  the  new  vector  fermions  are  {\em  still}  of  order
$\alpha_W$  instead  of order  $\alpha_S$,  as they  were for the $Z'$
coupling  \cite{Altarelli96}.  The introduction of new vector fermions
produces  new small FCNC  effects (b) at the  tree--level,  which are,
however,   not  in   disagreement   with  the   relevant   experiments
\cite{FCNC}.  The  limits,  imposed by FCNC  experiments,  exclude the
possibility  of low  energy  measurements  (c), like  $\nu N$  neutral
current scattering, as discussed, e.g., in \cite{McFarland96}  for the
case of the $Z'$.  

We employ in this  Letter  the  effect of the mixing  between  the new
quark vector  triplet and the SM quark flavours $q$ on the  production
of $q\bar{q}$  pairs at the LEP2  collider.  We do not claim that
that our $q\bar{q}$  studies are the unique  testing  ground:  further
studies must follow to find the best {\em signal/background} ratio for
this new vector fermions model.

To give a brief  outline of this  Letter:  we first give some  details
about the model we employ.  In Section~3  we fit the model  parameters
to recent values of $R_b$ and $R_c$ and present numerical  results for
various  $q\bar{q}$ cross sections at LEP2 energies.  We conclude with
some critical remarks  concerning the  detectability  of our predicted
effects in Section~4.

\section{Short Description of the Model}

We shall be very  cursory in the  description  of this model as it was
originally    motivated   and   discussed   in    Ref.~\cite{Chang96}.
Introducing the following vector quark triplet
\beq
X_{L,R}^{(Q,T^3)} = \left( \begin{array}{l}
x_1^{(5/3,+1)}\\
x_2^{(2/3,0)}\\
x_3^{(-1/3,-1)} \end{array} \right)_{L,R},
\eeq
with mass $M_X$ and the stated  quantum  numbers for charge  ($Q$) and
third  component  of weak  isospin  ($T^3$),  we  allow  a  mixing  of
$(x_3)_L$ with the $d$--type quarks $(d,s,b)_L$ and $(x_2)_L$ with the
$u$--type  quarks  $(u,c,t)_L$.  In the  context  of this  model it is
expected  that the  $M_X$ is much  bigger  than the  top--quark  mass.
However, we are exclusively  interested in the mixing effects this new
quark triplet shows with the quark  flavours of the SM.  The mixing of
the left--handed components are proportional to $1/M_X$, while for the
right--handed  couplings a $1/M_X^2$ dependence can be found, assuming
a large $M_X$ approximation, and therefore the latter can be neglected
(cf.     Refs.~\cite{Chang96,Branco86}).     The      neutral--current
Lagrangian, including $\gamma$ and $Z$ exchange, reads
\beq
{\cal{L}}_{\rm NC} = eJ^{\rm em}_\mu A^\mu + 
\sqrt{2}\left( \frac{G_F}{\sqrt{2}} \right)^{1/2}M_{Z}
\sum\limits_q \overline{\Psi}_q\gamma_\mu
(v_q - a_q\gamma_5)\Psi_q Z^{\mu},
\eeq
where the $\Psi_q$ are meant to be {\it  gauge--eigenstates}.  One can
immediately  see that taking into  account the unitary  transformation
matrix  that  shuffles  {\it  mass--}  into {\it  gauge--}eigenstates,
$\Psi_{qL,R}^{\rm gauge} \rightarrow  U_{L,R}\Psi_{qL,R}^{\rm  mass},$
modifies    the    isospin    matrices     $T^{3L,R}_q     \rightarrow
\overline{U}_{L,R}T^{3L,R}_q   U_{L,R}.$   As  a  result,  we  finally
conclude that the vector and axial--vector couplings defined by
\begin{eqnarray}
v_q &=& g_q^L + g_q^R = T_q^{3L} + T_q^{3R} - 2Q_q \sin^2\Theta_W, \\
a_q &=& g_q^L - g_q^R = T_q^{3L} - T_q^{3R},
\end{eqnarray}
are directly  influenced  by the presence of this new vector  triplet.
Consider, e.g., the new mass matrix $M_D$ for $d$--type quarks
\beq
M_D = \left( \begin{array}{cccc}
m_d & 0 & 0 & J_1 \\
0 & m_s & 0 & J_2 \\
0 & 0 & m_b & J_3 \\
0 & 0 & 0 & M_X \end{array} \right) = \left( \begin{array}{cc}
\widetilde{M}_D & J \\
0 & M_X \end{array} \right),
\eeq
where the matrix elements $J_i$ measure the relative {\em strength} of
the mixing between $(x_3)_L$ and the corresponding  $d$--type  quarks.
$U_L$  diagonalises  the matrix product  $M_DM_D^{\dagger}$.  Assuming
the most general form
\beq
U_L = \left( \begin{array}{cc} K & R \\
S & T  \end{array} \right),
\eeq
we find in the large $M_X$ approximation \cite{Branco86}
\beq
R = \frac{1}{M_X}J, \qquad S = -\frac{1}{M_X}J^{\dagger}K,
\eeq
with $K$ being the  $3\times3$  unity matrix.  

In complete  analogy we can study the $u$--type quark mixing  mediated
by the  matrix  $U'_L$  and  it can  be  shown   that  $S'=SK'$,  with
$K'\approx  V_{\rm  CKM}^{\dagger}$,  where  $V_{\rm CKM}$ denotes the
standard Cabibbo--Kobayashi--Maskawa matrix.

Consequently we derive for the new vector and axial--vector  couplings
in the case of the $u$-- and $d$--type quarks
\begin{eqnarray}
\widetilde{v}_u &=& v_u + (T^{3L}_{x_2}-T^{3L}_{u})|U'_{x_2u}|^2_L, 
\quad 
\widetilde{a}_u = a_u + (T^{3L}_{x_2}-T^{3L}_{u})|U'_{x_2u}|^2_L,\\
\widetilde{v}_d &=& v_d + (T^{3L}_{x_3}-T^{3L}_{d})|U_{x_3d}|^2_L, 
\quad 
\widetilde{a}_d = a_b+ (T^{3L}_{x_3}-T^{3L}_{d})|U_{x_3d}|^2_L. 
\end{eqnarray}
This explains the former special choice for the isospin  components of
the  vector  triplet.  As  $\Gamma(e^+e^-\rightarrow  q\bar{q})\propto
(\widetilde{v}_q^2 + \widetilde{a}_q^2)$, $T^{3L}_{x_2}=0$ will reduce
the hadronic width of the $u$--type quarks, whereas  $T^{3L}_{x_3}=-1$
enhances  the hadronic  width of the  $d$--type  quarks and  therefore
lower the SM value of $R_c$ and simultaneously  raise $R_b$.  In terms
of the above  formalism, the modified  couplings of, e.g., the $b$ and
$c$ quarks read now
\begin{eqnarray}
\widetilde{v}_c &=& v_c - \frac{1}{2}|S'_2|^2, \quad 
\widetilde{a}_c = a_c - \frac{1}{2}|S'_2|^2, \\
\widetilde{v}_b &=& v_b - \frac{1}{2}|S_3|^2, \quad 
\widetilde{a}_b = a_b - \frac{1}{2}|S_3|^2, 
\end{eqnarray}
where $S$ and $S'$ are to be identified  with the matrix elements used
in the  definition  of $U_L$ (cf.  Eq.(6)) and $U'_L$.  Thus the model
depends on 6 parameters ($S_{1,2,3}, S'_{1,2,3}$).  But because of the
relation $S'=SV_{\rm  CKM}^{\dagger}$, it is sufficient to constrain 3
parameters in order to calculate the full set.  The input to constrain
$S'_2$  and  $S_3$  will  be  $R_c$  and  $R_b$,  respectively.  After
determining   all   modified   vector  and   axial--vector   couplings
$\widetilde{v}_q$  and  $\widetilde{a}_q$,  we  proceed  to answer the
question of how the total $q\bar{q}$ cross section at LEP2 is affected
by the presence of the additional vector fermions.

The    total    cross    section    for   the    subprocess    $e^+e^-
\stackrel{Z,\gamma}{\rightarrow}  q\bar{q}$,  where $q$  stands  for
{\it one} quark flavour, via $Z$ and $\gamma$ exchange, reads in the
finite--quark--mass Born approximation
\begin{eqnarray}
\sigma^0(e^+e^-\stackrel{Z,\gamma}{\rightarrow} q\bar{q}) &=&
\frac{\beta}{2}(3-\beta^2)\sigma^0_V + 
\beta^3\sigma^0_A, \\
\sigma^0_V &=& \frac{4\pi\alpha^2}{s} Q_e^2Q_q^2 + 
\frac{4\alpha}{\sqrt{2}}G_FQ_eQ_qv_e\widetilde{v}_q
\frac{M_{Z}^2(s-M_{Z}^2)}{(s-M_{Z}^2)^2+M_{Z}^2\Gamma_Z^2} 
\nonumber \\ 
&+& \frac{G_F^2}{2\pi}\widetilde{v}_q^2(a_e^2+v_e^2)
\frac{sM_{Z}^4}{(s-M_{Z}^2)^2+M_{Z}^2\Gamma_Z^2}, \\
\sigma^0_A &=&  \frac{G_F^2}{2\pi}\widetilde{a}_q^2(a_e^2+v_e^2)
\frac{sM_{Z}^4}{(s-M_{Z}^2)^2+M_{Z}^2\Gamma_Z^2},
\end{eqnarray}
with  the  quark  velocity  $\beta=\sqrt{1-4\frac{m_q^2}{s}}$  and the
electron couplings $v_e = -1/2 + 2\sin\Theta_W$ and $a_e = -1/2$.  For
further details we refer to, e.g., \cite{Jersak81}.  We implemented in
our calculations the following corrections:

\begin{itemize}  
\item
{QCD  ${\cal{O}}(\alpha_S)$  corrections due to real and virtual gluon
emission \cite{Jersak81},}
\item
{universal    QED     ${\cal{O}}(\alpha)$     corrections     $3\alpha
Q_q^2/(4\pi)$,}
\item
{initial state radiation up to  ${\cal{O}}(\alpha^2)$  in QED and soft
photon exponentiation \cite{Berends88},}
\item
{universal   electroweak   corrections  due  to  $tt$  and  $tb$  loop
corrections to the $Z$ propagator and $\gamma$--$Z$  mixing as well as
vertex  corrections for $b$ quarks due to virtual $t$ quark  exchanges
\cite{Akhundov88}. The  box--graph   contributions   turn  out  to  be
unimportant for our purposes \cite{LEP96}.}
\end{itemize}

\section{Numerical Results}

For  our  numerical  analysis  we  used  the  {\it  on--shell}  scheme
addressing the following  electroweak data:  $\alpha(M_{Z}) = 1/128.8,
M_{Z} =  91.188$~GeV,  $M_W =  80.33$~GeV  with  the  strong  coupling
constant $\alpha_S(M_{Z}) = 0.123.$ Furthermore we fixed the top quark
mass to be $m_t = 175$~GeV.  For the CKM--matrix  elements we used the
averaged values given in Ref.~\cite{PDG96}.

Starting  from the  latest  set of  values  for  $R_b$  and $R_c$  (as
discussed  above), we first  performed a general fit of the  parameter
space, including $|S'_2|^2$ and $|S_3|^2$, as they directly govern the
values for $\Gamma_{c\bar{c}}$ and  $\Gamma_{b\bar{b}}$.  Furthermore,
following  Ref.~\cite{Chang96},  we set  $|S_1|^2\simeq  0$.  This  is
invoked  by  taking  into  account  the  FCNC  of the  process  $K^0_L
\rightarrow  \mu^+\mu^-$, which does not support any sizeable $d$--$s$
quark mixing.  The missing values  $|S'_1|^2,  |S'_3|^2$ and $|S_2|^2$
are   then   calculated,    consistent   the   relation    $S'=SV_{\rm
CKM}^{\dagger}$.  Although we found a weak  dependence  of the various
widths ($\Gamma_{Z},  \Gamma^{\rm  had}_{Z}$) on the input parameters,
the cross sections remain quite  insensitive to the widths compared to
the modified  coupling--dependence,  as can easily be deduced from the
formal expression of the total cross section in Eqs.~(12)--(14).

Figure~(1)  shows our fitted values for $|S'_2|^2$  and  $|S_3|^2$, in
particular the edges of the $1\sigma$ (68.3\%  confidence level of the
normal distribution) and the $2\sigma$ (95.4\% c.l.)  regions of $R_b$
and $R_c$.  We find  $|S'_2|^2 = 0.01245$ and $|S_3|^2 = 0.00922$  for
$1\sigma$  (corresponding  to $R_b = 0.2189$  and $R_c = 0.1659$)  and
$|S'_2|^2 = 0.02528$ and  $|S_3|^2 = 0.01284$ for the  $2\sigma$  case
($R_b = 0.2200$ and $R_c = 0.1603$).

With  these two sets of  parameters  deduced  from  $R_b$ and $R_c$ we
first   give   predictions   for   the   subprocess   cross   sections
$\sigma(e^+e^-\rightarrow  s\bar{s},c\bar{c},b\bar{b})$,  as they turn
out to give  the  most  significant  signal.  Specifically,  Fig.~(2a)
shows the  $c\bar{c}$  production  cross  section as a function of the
centre--of--mass  energy.  The  contribution of the additional  vector
fermions to this cross section is negative, as it is for all $u$--type
quarks,  which can  easily be  checked  from  Eq.~(8),  as the  former
motivation is to {\it  decrease} the SM value of $R_c$.  The result is
very  sensitive  to the  values  of $R_b$ and  $R_c$,  as,  e.g.,  the
$1\sigma$  input and the $2\sigma$ input differ by a factor of roughly
2.  Moreover it can be observed  that in the energy region of the LEP2
collider    ($160~$GeV--$190$~GeV),   the   contribution   is   nearly
insensitive  to  $\sqrt{s}$.  The observed  gaps in all figures  which
appear  around  the $Z$  mass  $M_{Z}  =  91.188$~GeV,  are due to the
resonant behaviour of the total cross section.

\begin{table}[h]

\begin{center}
\begin{tabular}{|cl|c|c||c|c|}\hline
 
&  &   \multicolumn{2}{|c||}{{\shift   $1\sigma$   (68.3\%  c.l.)}}  &
\multicolumn{2}{c|}{$2\sigma$ (95.4\% c.l.)}\\ \cline{3-6} & & {\shift
$\sigma_{q\bar{q}}$}  \lbrack pb  \rbrack &  $\delta\sigma_q/\sigma_q$
\lbrack  \%  \rbrack  &  $\sigma_{q\bar{q}}$   \lbrack  pb  \rbrack  &
$\delta\sigma_q/\sigma_q$     \lbrack    \%    \rbrack    \\    \hline
$\sqrt{s}=$&${\shift  c\bar{c}}$  & 36.69 & $-5.31$ & 34.71 & $-10.44$
\\ \cline{2-6}  $175$~GeV&${\shift b\bar{b}}$ & 12.75 & +3.83 &
12.93 & +5.31  \\  \hline  $\sqrt{s}=$&${\shift  c\bar{c}}$  & 29.75 &
$-5.26$  &  28.15  &  $-10.35$   \\   \cline{2-6}   $190$~GeV&${\shift
b\bar{b}}$ & 10.16 & +3.82 & 10.31 & +5.30 \\ \hline \end{tabular}

\caption[]{   The   numerical   values   for    $\sigma_{q\bar{q}}   =
\sigma(e^+e^-\rightarrow  q\bar{q})$ and  $\delta\sigma_q/\sigma_q$ in
the vector  fermion  model, for $q$ being either a $c$ or a $b$ quark.
The  calculations  were performed for two typical values of $\sqrt{s}$
at LEP2, and for the two  confidence  levels as discussed in the text.
}

\end{center} 
\end{table}

All  the  argumentation  drawn  from  Fig.~(2a)  also  holds  for  the
discussion of the $s\bar{s}$ and $b\bar{b}$ cross sections,  presented
in   Figs.~(2b)--(2c).  The  main   difference   is  that  the   total
contribution  of the additional  vector fermions is {\it positive} for
$d$--type   quarks.  Again  we  refer  to  Eq.~(9).  Even  though  the
absolute  value  of  the  $s\bar{s}$  contribution  is  comparable  to
Fig.~(2a),  there is no hope to isolate  this cross  section in a LEP2
measurement.  The   $b\bar{b}$   cross   section,   however,  will  be
measurable  and  according  to  Ref.~\cite{Leike96}  the  experimental
uncertainty  in  $\sigma_{b\bar{b}}$  is  reported  to  be  2.5\%  for
$\sqrt{s}=190$~GeV and an assumed luminosity of $500$~pb$^{-1}$, which
allows for a clear signal.  Although former studies at the LEP2 showed
a general  lower  accuracy  for the tagging of  $c\bar{c}$  events, it
still  might be  sufficient  to detect  our  calculated  5\% effect in
$\sigma_{c\bar{c}}$  as shown in Fig.~(2a).  All numerical results are
summarised in Table~1 for two fixed LEP2 centre--of--mass  energies of
$\sqrt{s}=175$~GeV and 190~GeV.

Finally we present in Fig.~(3) the total cross  sections for $d$-- and
$u$--type  quarks.  Again, we can see the overall tendency that the SM
cross  section  is being  lowered  under the  presence  of the  vector
fermions for $u$--type  quarks,  whereas we find a proper  enhancement
for  $d$--type  quarks,  which is the  characteristic  feature of this
model.  We can not expect a tagging of these individual cross sections
at this level of accuracy at LEP2, but for reasons of completeness  we
want to mention it at this stage, especially to demonstrate that there
will be no signal to be expected in the {\em total} cross  section, as
individual  subprocess  contributions  will  cancel  each  other to an
almost zero level.

\section{Remarks and Conclusion}

In this Letter we studied the impact of a model with additional vector
fermions at the LEP2 collider.  We made  predictions for various quark
production  cross sections and discussed their possible  detectability
based   on   a   recent    phenomenological    analysis    given    in
Ref.~\cite{Leike96}.  However,  there is probably no evidence  for new
physics at this stage,  especially after the values of $R_b$ and $R_c$
are  approaching  the SM  predictions,  although,  one can argue  that
exploiting the idea of additional  vector  fermions at energies beyond
the $Z$ pole is of considerable interest.  A remarkable feature of the
model  we were  dealing  with  throughout  our  studies  is that it is
anomaly  free, in contrast  to  alternative  ideas  emerging  from the
reported ``crisis", and therefore seems to be more physically sound.

Even though our predictions yield relatively small effects, within the
scope of accuracy at LEP2 they might be  measureable.  Present studies
concerning  statistical  and  systematic  errors  at LEP2 are  already
underway.  Especially taking into account that the overall corrections
are different in sign for $u$-- and $d$--type quarks, the simultaneous
tagging of  $c\bar{c}$  and  $b\bar{b}$  pairs  might  further  reduce
statistical  errors.  We  believe  that this will   yield, as a unique
feature, a {\em clear} signal due to the existing splitting around the
SM predictions, as we have already discussed in Section~3.

As a final  remark we want to mention  that we also checked a possible
influence on the  forward--backward  asymmetries  $A_{\rm  FB}^q$.  At
LEP1  energies the modified  asymmetries  are within the  experimental
errors,  as  already  discussed  in  Ref.~\cite{Chang96}  and by Ma in
Ref.~\cite{VFModels96}.  The  accuracy  at LEP2 for the  corresponding
measurements  is expected to be even lower, such that it is  difficult
to draw any conclusions from forward--backward asymmetry measurements.

\section*{Acknowledgements}

We  want to  thank  James  Stirling  for a  careful  reading  of  this
manuscript and critical remarks.  The work of V.C.S.  was supported by
the  European  Union  under the Human  Capital  and  Mobility  Network
Program   CHRX--CT93--0319.  M.H.  gratefully  acknowledges  financial
support in the form of a ``DAAD--Doktorandenstipendium'' HSP--II/AUFE.


\begin{figure}[b]             
\unitlength1cm            
\begin{center}
\begin{picture}(13.2,16)            
\makebox[13.2cm]{\epsfxsize=13.2cm
\epsfysize=14cm 
\epsffile{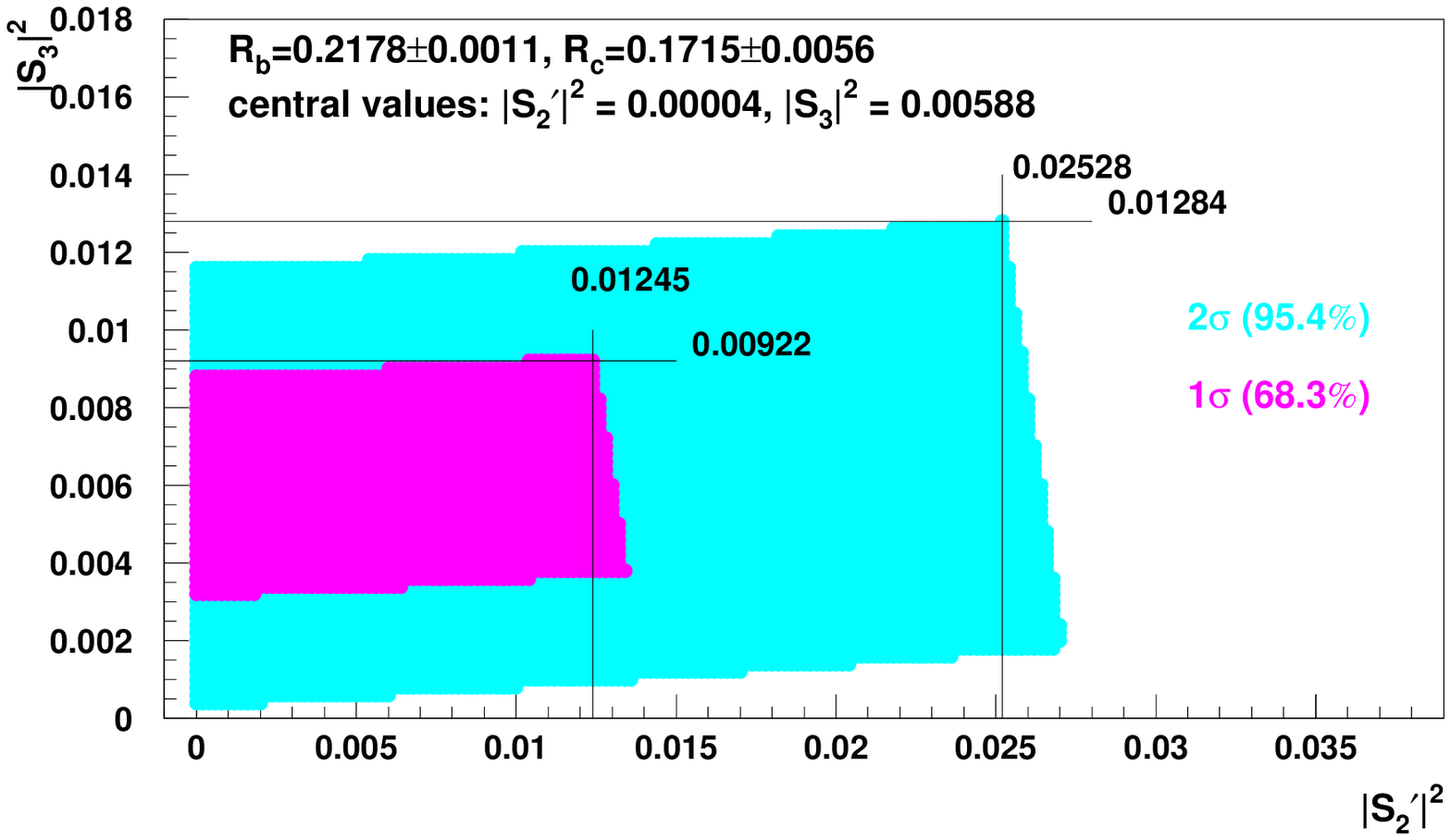} } 
\end{picture}
\end{center}

\vspace{-5cm}

\caption[]{We  show the allowed  regions in the  $|S_2'|^2$--$|S_3|^2$
plane (as discussed in the text) obtained by fitting the recent $R_b$,
$R_c$ values \cite{Blondel96}.  We present the $1\sigma$ (68.3\% c.l.)
and $2\sigma$ (95.4\% c.l.)  regions, from which we read off our input
parameters $|S_2'|^2$ and $|S_3|^2$.  }

\end{figure}


\begin{figure}[b]             
\unitlength1cm            
\begin{center}
\begin{picture}(13.2,16)            
\makebox[13.2cm]{\epsfxsize=13.2cm
\epsfysize=14cm 
\epsffile{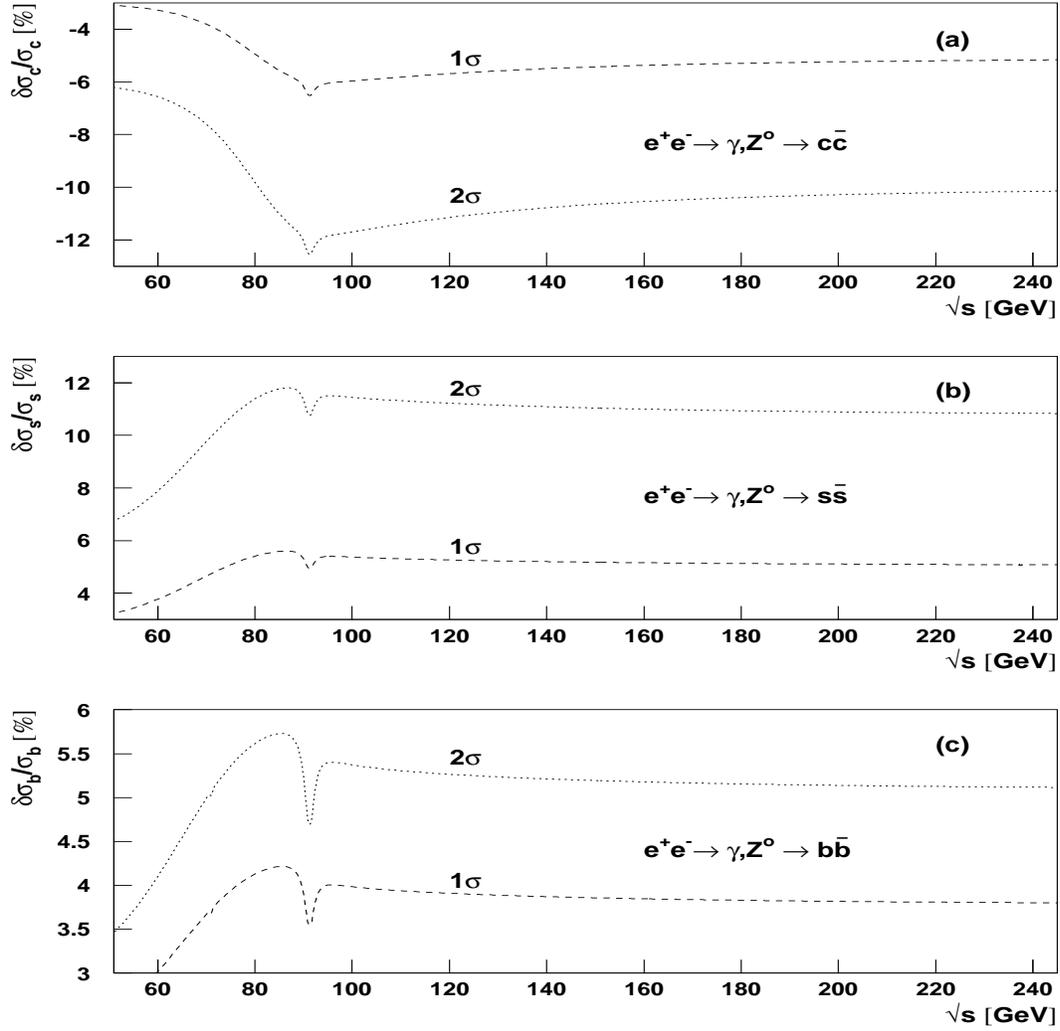} } 
\end{picture}
\end{center}

\caption[]{We present the relative differences between the predictions
of the vector fermion (VF) model and the SM  ($\delta\sigma_q/\sigma_q
:=     (\sigma_{q\bar{q}}^{\rm     VF}    -     \sigma_{q\bar{q}}^{\rm
SM})/\sigma_{q\bar{q}}^{\rm  SM}$)  in  per  cent  as  a  function  of
$\sqrt{s}$, for three different flavours.}

\end{figure}


\begin{figure}[b]             
\unitlength1cm            
\begin{center}
\begin{picture}(13.2,16)            
\makebox[13.2cm]{\epsfxsize=13.2cm
\epsfysize=14cm 
\epsffile{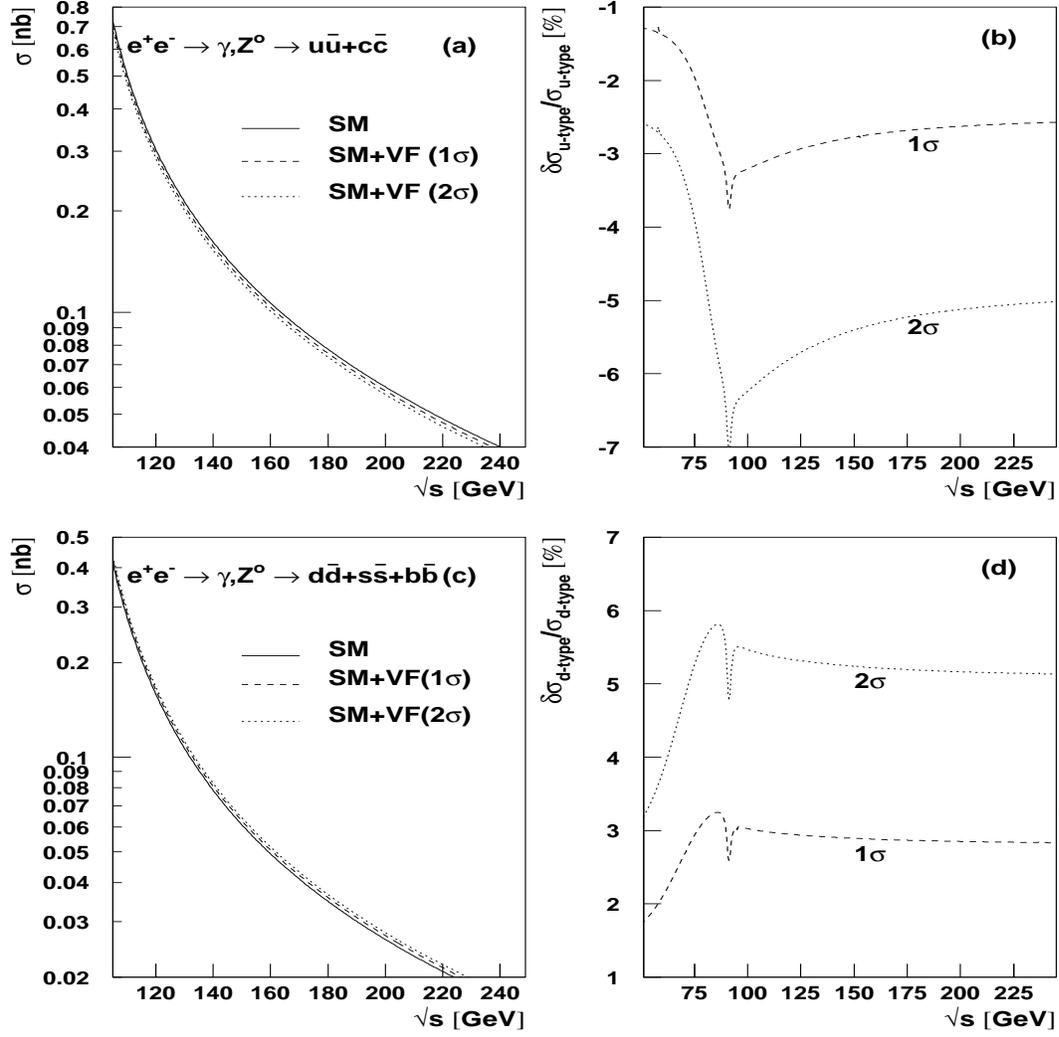} } 
\end{picture}
\end{center}

\caption[]{The  total cross  section for $u$-- (a) and  $d$--type  (c)
quarks  in the SM  calculation  and the  two  vector  fermion  fits as
discussed in the text,  including the relative  changes (b) and (d) as
already shown in Fig.~2.  }

\end{figure}

\end{document}